\documentclass[12pt]{iopart}

\begin{document}

\title{Boundary resistance in magnetic multilayers}

\author{B P Vodopyanov$^{1}$ and L R Tagirov$^{1,2}$\footnote[3]{To
whom correspondence should be addressed (Lenar.Tagirov@ksu.ru)}}

\address{$^{1}$Kazan Physical-Technical Institute of RAS, 420029 Kazan, Russia}
\address{$^{2}$Kazan State University, 420008 Kazan, Russia}

\begin{abstract}
Quasiclassical boundary conditions for electrochemical potentials at the
interface between diffusive ferromagnetic and non-magnetic metals are
derived for the first time. An expression for the boundary resistance
accurately accounts for the momentum conservation law as well as essential
gradients of the chemical potentials. Conditions are established at which
spin-asymmetry of the boundary resistance has positive or negative sign.
Dependence of the spin asymmetry and the absolute value of the boundary
resistance on the exchange splitting of the conduction band opens up new 
possibility to estimate spin polarization of the conduction band of
ferromagnetic metals. Consistency of the theory is checked on existing
experimental data.
\end{abstract}

\pacs{74.50. + r, 74.80. Dm, 74.62.-c}

\submitto{Published in J. Phys.: Condens. Matter. \textbf{18}, 1545-1552 (2006)}

\maketitle

\section{Introduction}

Discovery of giant magnetoresistance (MR) in magnetic multilayers \cite
{Baibich, Binash}, which consist of alternating ferromagnetic metal (F) 
and normal metal (N) layers, stimulated intensive studies of spin-polarized 
transport through layered magnetic structures \cite{Gijs, Bass,Zutic}. 
In a paper \cite{Valet} the semiclassical theory of MR in magnetic 
multilayers has been developed for geometry, in which the electric current 
is perpendicular to the layers (CPP geometry). It has been shown that when 
thickness of the metals in a stack is small compared with the spin diffusion 
length, the magnetoresistance of multilayers can be calculated within the 
two-current series-resistor model \cite{Zhang91, Bauer, Lee}. In this case 
MR is expressed via the resistance of the F/N interface (boundary resistance), 
$r_{\uparrow (\downarrow )}=2[1\mp \gamma _{F/N}]R_{F/N}^{\ast }$, and the bulk
specific resistance of the ferromagnetic layer, $\rho _{\uparrow (\downarrow
)}=2[1\mp \beta _{F}]\rho _{F}^{\ast }$ \cite{Valet}. In these expressions 
$\gamma _{F/N}$ and $\beta _{F}$ are parameters of spin asymmetry of the
boundary and bulk scattering resistances. It follows from works \cite{Valet,
Rashba} that their linear combination, $\beta _{F1}\rho _{F1}^{\ast
}t_{F1}+\gamma _{1}R_{F/N}^{\ast}$, determines sign of MR in layered 
$[F1/N/F2/N]\times n$ structures ($t_{F}$ is the thickness of the
ferromagnetic layers). In works \cite{Vouille,Vouille1} positive (inverse)
MR due to negative value of $\gamma _{F/N}$ in $(F1/Cr/F2/Cr)$ multilayers
has been observed for the first time.

Theoretical calculations of the boundary resistance \cite{Schep, Stiles, 
Xia} established the strong influence of the spin-dependent band structure of
ferromagnetic metals on magnetoresistance. Using the approach of Ref. \cite
{Schep} and transmission probabilities through the F/N interface, calculated 
earlier from first principles \cite{Stiles2}, Stiles and Penn \cite{Stiles} 
obtained from numerical calculations a negative value of $\gamma_{F/N}$ for the 
$Fe/Cr$ interface, and a positive $\gamma _{F/N}$ for iron-group ferromagnet - 
noble metal interfaces. However, conditions on the parameters of contacting 
materials resulting in negative or positive values of spin-asymmetry of the
boundary resistance were not discussed in the above papers.

In this paper we derive for the first time quasiclassical boundary
conditions for electrochemical potentials of diffusive ferromagnetic and
normal metals, which can be used for solution of a wide class of problems in
spintronics. Our expression for the boundary resistance accurately accounts
for the momentum conservation law as well as essential gradients of the
chemical potentials. We establish conditions on parameters of the contacting
metals, at which spin asymmetry of the boundary resistance has a positive or
negative sign. Dependence of the spin asymmetry and the boundary resistance
on the exchange splitting of the conduction band offers one more way to estimate
spin polarization of conduction band of ferromagnetic metals. We give an
example of such an estimation.

\section{Boundary conditions for electrochemical potentials}

We derive boundary conditions for electrochemical potentials of diffusive
metals using the quasiclassical Green functions technique. The "diffusive
ferromagnetic metal" is a metal in which spin splitting of the conduction band
is small compared with the momentum relaxation rate for conduction
electrons. Let us consider that axis $x$ is perpendicular to the $F/N$
boundary, and neglect reversal of the electron spin upon transmission
through the interface. Then, for each of the metals equations for the Green
functions $g_{\alpha }(\mathbf{n},x,\rho ,t)$ read \cite{Zaitsev}:
\begin{equation}
v_{x,\alpha }\frac{\partial g_{a,\alpha }}{\partial x}+\mathbf{v}_{\parallel
}\frac{\partial g_{s,\alpha }}{\partial \mathbf{\rho }}+\frac{1}{\tau
_{\alpha }}\left( g_{s,\alpha }-\overline{g}_{s,\alpha }\right) =0,
\label{eq1} 
\end{equation}
\begin{equation}
v_{x,\alpha }\frac{\partial g_{s,\alpha }}{\partial x}+\mathbf{v}_{\parallel
}\frac{\partial g_{a,\alpha }}{\partial \mathbf{\rho }}+\frac{g_{a,\alpha }}
{\tau _{\alpha }} =0.  \label{eq2}
\end{equation}
Here $\mathbf{n}=\mathbf{p}_{x,\,\alpha }/|\mathbf{p}_{\,\alpha }|$; 
$g_{s(a),\,\alpha }=1/2[g_{\,\alpha }(n_{x},x,\rho ,t)\pm g_{\,\alpha
}(-n_{x},x,\rho ,t)]$ is the single-particle quasiclassical Green function
symmetric (antisymmetric) with respect to a projection of the Fermi momentum 
$\mathbf{p}_{x,{\,\alpha }}$ on the axis $x$ ; $v_{x}$ is a projection of
the Fermi velocity on the axis $x$; $\alpha =(\uparrow ,\downarrow )$ is a
spin index, and $\mathbf{\rho }=(y,z)$ is a coordinate in a plane of the
contact. The bar above $g_{s,\alpha }$ means integration over the solid angle: 
$\overline{g}_{s,\,\alpha }=\oint d\Omega /2\pi \;g_{s,\alpha }$.

The boundary conditions to Eqs. (\ref{eq1}) and (\ref{eq2}) are as follows 
\cite{Zaitsev}: 
\begin{eqnarray}\label{eq3}
g_{\,a,\alpha}^{F}(0) = g_{a,\alpha}^{N}(0)=\left\{
g_{a,\alpha}(0 ),\,\,\,\,p_{\Vert}<p_{\,\alpha}^{\,F},p^{\,N}
\atop 0,\, \,\,\,\min
(p_{\alpha}^{F},p^{N})<p_{\Vert } \right. , 
\end{eqnarray}

\begin{equation}
2R_{\alpha \,}g_{a,\alpha }(0)=D_{\alpha }\left( g_{s,\alpha
}^{F}(0)-g_{s,\alpha }^{N}(0)\right) .  \label{eq4}
\end{equation}
In the equations (\ref{eq3}) and (\ref{eq4}) $p_{\alpha }^{F}$ and $p^{N}$ 
are the Fermi momenta in ferromagnetic and normal metals, respectively; 
$p_{\parallel }$ is a projection of a momentum on the plane of the contact; 
$D_{\alpha }$ and $R_{\alpha }=1-D_{\alpha }$ are the spin-dependent,
quantum-mechanical transmission and reflection coefficients. Boundary
conditions (\ref{eq3}), (\ref{eq4}) obey the specular reflection law:
\begin{equation}
p_{\parallel }=p_{\downarrow }^{F}\sin \theta _{\downarrow }=p_{\uparrow
}^{F}\sin \theta _{\uparrow }=p^{N}\sin \theta _{N}.  \label{eq5}
\end{equation}
The angles $\theta $ in (\ref{eq5}) are measured from the axis $x$, a range
of variation for the biggest one is $[0,\pi /2]$. The quasiclassical
equations (\ref{eq1}) and (\ref{eq2}), and the boundary conditions (\ref{eq3}
) and (\ref{eq4}) are formulated for a single electron trajectory determined
by the angles $\varphi $ and $\theta$.

Upon solution of the system of equations (\ref{eq1}) and (\ref{eq2}) we
shall consider that the ferromagnet is located to the left of the boundary 
$x=0$, and the normal metal - to the right ($x>0$), and that the functions 
$g_{s,\alpha }$ are homogeneous in the plane of the contact. Then, the 
system of equations (\ref{eq1}) and (\ref{eq2}) can be solved in the form of
integral equations for the functions $g_{a,\alpha }$ and $g_{s,\alpha }$ in
the energy representation, $g_{s,\alpha }(\varepsilon )=2\tanh (\varepsilon
/2T)+f_{s,\alpha }(\varepsilon )$:
\begin{equation}
f_{s,\alpha }^{\,N}(x) =g_{a,\alpha }^{N}(x)+\frac{1}{l_{x,\alpha }}{
\int\limits_{x}^{\infty }}d\xi e^{\frac{x-\xi }{l_{x,\alpha }}}\overline{f}
_{s,\alpha }^{N}(\xi ),  \label{eq6} 
\end{equation}
\begin{equation}
f_{s,\alpha }^{\ F}(x) =-g_{a,\alpha }^{F}(x)+\frac{1}{l_{x,\alpha }}{
\int\limits_{-\infty }^{x}}d\xi e^{\frac{\xi -x}{l_{x,\alpha }}}\overline{f}
_{s,\alpha }^{F}(\xi ).  \label{eq7}
\end{equation}
In a dirty metal the solid-angle averaged function $\overline{f}_{s,\alpha
}(\xi )$ obeys the diffusion equation with a decay length which is much longer
than the mean free path $l_{\alpha }$. Then, we expand $\overline{f}
_{s,\alpha}^{F(N)}(\xi )$ in the r.h.s. of Eqs. (\ref{eq6}) and (\ref{eq7}) 
near point $x$ and take out from the integrals independent on $\xi$ terms. 
Substituting the resulting expansions into the boundary condition 
(\ref{eq4}) we find:
\begin{equation}
2\,g_{a,\alpha }(0)= 
D_{\alpha }\left[ \left( 1-l_{x,\alpha }^{F}\frac{d}{dx}\right) \overline{f}
_{s,\alpha }^{\,F}(x)-\left( 1+l_{x,\alpha }^{N}\frac{d}{d\,x}\right) 
\overline{f}_{s,\alpha }^{\,N}(x)\right] _{x=0}.  \label{eq8}
\end{equation}
To formulate boundary conditions for the functions $\overline{f}_{s,\alpha
}^{\,F(N)}$ (which are, in fact, chemical potentials - see below) we use
a matching procedure proposed in Ref. \cite{Kup}. From equation 
(\ref{eq1}) it follows that for distances of the order of $l_{x,\,\alpha}$ 
from the interface
\begin{equation}
\overline{l_{x,\alpha }\frac{dg_{a,\alpha }}{dx}}=0.  \label{eq9}
\end{equation}
Hence,
\begin{equation}
\overline{l_{x,\alpha }\,g_{a,\alpha }}=C=const  \label{eq10}
\end{equation}
in each of the metals. Now we calculate, for example, $C^{F}$ using the
expression (\ref{eq8}) for $g_{a,\alpha }(x=0)$. Then, we calculate $C^{F}$
far from the interface using an approximate expression for $g_{a,\alpha
}^{F}(x)$,
\begin{equation}
g_{a,\alpha }^{\,F}(x)=-l_{x,\alpha }^{F}\,\frac{d\overline{f}_{s,\alpha
}^{F}(x)}{dx},  \label{eq11}
\end{equation}
which follows from equation (\ref{eq2}) after expansion of $g_{a,\alpha
}^{F}$ on Legendre polynomials. Equating values of the constant $C^{F}$
calculated in the two ways, and applying relationship between the averaged
Green function and the electrochemical potential, $\overline{f}
_{s,\alpha }=(2/\pi )\mu _{\alpha },$ we receive the boundary condition for 
the electrochemical potentials at the interface $x=0$:
\begin{equation}
l_{\alpha }^{F}\frac{d\mu _{\alpha }^{F}(0)}{dx}=\delta _{\alpha }\left( \mu
_{\alpha }^{N}(0)-\mu _{\alpha }^{F}(0)\right) ,  \label{eq12}
\end{equation}
where
\begin{eqnarray}
\delta _{\alpha } =\frac{\delta _{1,\alpha }}{1-\delta _{2,\alpha }},\quad
\delta _{1,\alpha }=\frac{3}{2}\int \frac{d\Omega _{F,\alpha }}{2\pi }\,\cos
(\theta _{F,\alpha })D_{\alpha }, \nonumber \\
\delta _{2,\alpha } =\frac{3}{2}\int \frac{d\Omega _{F,\alpha }}{2\pi }
\left[ x+\left( \frac{p_{\alpha }^{\,F}}{p^{N}}\right) ^{2}\cos (\theta _{N})
\right] xD_{\alpha },  \label{eq13} \\
x =\cos (\theta _{F,\alpha }),\ \ \,\,d\Omega _{F,\alpha }=\sin (\theta
_{F,\alpha })\,d\theta _{F,\alpha }\,d\varphi . \nonumber 
\end{eqnarray}
The limits of angular integration must satisfy specular reflection
conditions at the interface, Eq. (\ref{eq5}). When deriving Eq. (\ref{eq12})
we have used conservation of the current density at the interface, which
follows from Eq. (\ref{eq3}), 
\begin{equation}
j_{\alpha }^{F}(0)=\frac{\sigma _{\alpha }^{F}}{e}\frac{d\mu _{\alpha
}^{F}(0)}{dx}=\frac{\sigma _{\alpha }^{N}}{e}\frac{d\mu _{\alpha }^{N}(0)}{dx
}=j_{\alpha }^{N}(0),  \label{eq14}
\end{equation}
where $\sigma _{\alpha }^{\,F}$ and $\sigma _{\alpha }^{\,N}$ are the bulk,
spin-channel conductivities of the metals:
\begin{equation}
\sigma _{\alpha }^{\,F(N)}=\frac{e^{2}(p_{\alpha }^{F(N)})^{2}l_{\alpha
}^{F(N)}}{6\pi ^{2}}.  \label{eq15}
\end{equation}
Equation (\ref{eq14}) is actually the second, complementary to Eq. (\ref
{eq12}), boundary condition for the semiclassical description of the
spin-polarized transport in magnetic multilayers in terms of electrochemical
potential.

\section{Resistance of the interface}

The derivative from the electrochemical potential in Eq. (\ref{eq12}) can
be expressed in terms of the density of current (\ref{eq14}), and we find
the spin-dependent resistance of the interface $r_{\alpha }$:
\begin{equation}
\mu _{\alpha }^{N}(0)-\mu _{\alpha }^{F}(0)=er_{\alpha }\,j_{\alpha },
\label{eq19}
\end{equation}
\begin{equation}
r_{\alpha }=\frac{6\pi ^{2}}{e^{2}(p_{\alpha }^{F})^{2}A}\frac{1-\delta
_{2,\alpha }}{\delta _{1,\alpha }},  \label{eq20}
\end{equation}
where $A$ is the area of the contact. It follows from equation (\ref
{eq20}) that, in the quasiclassical approach at specular reflection from 
the interface, the boundary resistance between ferromagnetic and normal 
metals is determined only by the Fermi momenta of the contacting metals 
and coefficient of transmission through the interface.

Experimental data are given for the spin asymmetry of boundary resistance, 
$\gamma _{F/N}$, and for the renormalized resistance of the interface, 
$AR_{F/N}^{\ast }$, determined as follows:
\begin{equation}
\gamma _{F/N}=\frac{r_{\downarrow }-r_{\uparrow }}{r_{\uparrow
}+r_{\downarrow }},\,\,\,\,\,\,AR_{F/N}^{\ast }=\frac{A}{4}(r_{\downarrow
}+r_{\uparrow }).  \label{eq21}
\end{equation}
To calculate dependence of $\gamma _{F/N}$ and $AR_{F/N}^{\ast }$ on the 
Fermi momentum of the non-magnetic metal for various values of the 
ferromagnet conduction band polarization we have used the 
Fermi-momentum-mismatch model for the transmission coefficient: 
$D_{\alpha}=4v_{x,\alpha}^{N}v_{x,\alpha}^{F}/[(v_{x,\alpha}^{N})^2+
(v_{x,\alpha}^{F})^2]$. Results are presented in Figures 1 
($p_{\uparrow }^{F}>p_{\downarrow}^{F}>p^{N}$), 
2 ($p_{\uparrow }^{F}>p^{N}>p_{\downarrow }^{F}$) and 3 
($p^{N}>p_{\uparrow }^{F}>p_{\downarrow }^{F}$). From our calculations it
follows that for a non-magnetic metal with a low density of conduction 
electrons (small value of the Fermi momentum $p^{N}$) the spin asymmetry of 
the boundary resistance $\gamma _{F/N}$ is always negative (Fig. 1). On the
contrary, for a non-magnetic metal with a high density of conduction 
electrons the spin asymmetry of the boundary resistance is always positive 
(Fig. 3). In an intermediate situation $\gamma_{F/N}$ can be negative 
as well as positive (see Fig. 2). To attain the maximum amplitude 
of negative magnetoresistance in $F/N$ multilayers the spin 
asymmetry of the boundary resistance $\gamma _{F/N}$
and the asymmetry of the bulk resistance $\beta _{F}$ should be both
positive and close to unity. According to our calculations, the Fermi
momentum of the non-magnetic metal should be as far as possible close to the
Fermi momentum of the majority subband of ferromagnetic metal (Fig. 2, 
$p^{N}/p_{\uparrow }^{F}\rightarrow 1.0$ and Fig. 3, $p_{\uparrow
}^{F}/p^{N}\rightarrow 1.0$). The spin asymmetry of the bulk resistance can 
be adjusted by the type and concentration of impurities in the ferromagnetic 
metal \cite{F-C}. Clearly, similar arguments can be applied to the 
opposite case of negative values of $\gamma _{F/N}$ and $\beta _{F}$, which 
will result in positive magnetoresistance in multilayers of alternating
ferromagnetic and non-magnetic metals. However, a negative asymmetry of
the bulk resistance is met less often \cite{F-C}. Competition of opposite
in a sign asymmetries of boundary and bulk resistances can result in
negative or positive magnetoresistance depending on the choice of materials
and thickness of the ferromagnetic layers \cite{Vouille1}.

\section{Discussion of experiments}

Experiments on CPP transport in multilayers are very complicated because 
the resistance of a stack of layers of nanometer thickness is very small
(order of $f\Omega $ m$^{2}$). Nevertheless, available experimental data
(see reviews \cite{Bass, Bass2} and references in them, and also works \cite
{Vouille1, Bass3,Lee2}) allow to test the internal consistency of the theory.
Multilayers of the iron-group ferromagnetic metals with noble metals most 
likely belong to case 3 ($p^{N}>p_{\uparrow }^{F}>p_{\downarrow }^{F})$, 
and parameter of the spin asymmetry $\gamma_{F/N}$ is positive \cite
{Bass,Vouille1,Bass2,Bass3,Lee2}. For example, $\gamma_{Co/Cu}\simeq 0.77$ 
\cite{Bass}. Then, intersection of the horizontal dash line $\gamma
_{Co/Cu}\simeq 0.77$ in the top field of Fig. 3 with the curve $\gamma
_{F/N}(p^{N})$, corresponding to $\delta =0.6$, gives $p_{\uparrow
}^{F}/p^{N}\simeq 0.7$. Accepting $p_{\uparrow }^{F}=1.0\mathring{A}^{-1}$
as a trial value for the Fermi momentum of the majority subband of cobalt we
receive $p^{Cu}\simeq 1.41\mathring{A}^{-1}$, which fits fairly good the
free-electron-model value for copper, $p_{FEM}^{Cu}\simeq 1.36\mathring{A}
^{-1}$ \cite{Ashk-Merm}. There are data for silver as the non-magnetic spacer: 
$\gamma_{Co/Ag}\simeq 0.85$ \cite{Lee2}. In a similar way, we obtain from
Fig. 3, $p^{Ag}\simeq 1.22\mathring{A}^{-1}$, which fits well the
free-electron-model value $p_{FEM}^{Ag}\simeq 1.20\mathring{A}^{-1}$ \cite
{Ashk-Merm}.

Let us look now at consistency of the theory with boundary resistance data.
For the combination $Co/Cu$, $AR_{Co/Cu}^{\ast }(\exp )\simeq 0.51\,f\Omega
\,m^{2}$ \cite{Bass}. Continuing the vertical dash line for copper in Fig. 3
into the bottom field till intersection with the curve $AR_{F/N}^{\ast
}(p^{N})$, corresponding to $\delta =0.6$, we obtain $AR_{Co/Cu}^{\ast }
(\mathrm{theor})\simeq 0.74\,f\Omega \,m^{2}$. A similar procedure gives
for silver $AR_{Co/Ag}^{\ast }(\mathrm{theor})\simeq 0.69\,f\Omega \,m^{2}$
(compare with $AR_{Co/Ag}^{\ast }(\exp )\simeq 0.56\,f\Omega \,m^{2}$ \cite
{Lee2,Vouille1}). It is worthy to notice that the theory reproduces fairly 
good closeness of the boundary resistances of the Co/Cu and Co/Ag 
interfaces.

For $Co/Cr$ multilayers the asymmetry of the boundary resistance is
negative, $\gamma_{Co/Cr}\simeq - 0.24$ \cite{Vouille1}. This value is
admissible for cases 1 (Fig.1, top field) and 2 (Fig. 2, top field).
Without details, we conclude that the first case does not match the expected
value of the conduction band polarization parameter for Co $\delta \sim
0.6\pm 0.1$ as well as results in about two orders in magnitude higher
boundary resistance. The second case (see dash lines in Fig. 2) results in
a Fermi momentum $p^{Cr}\sim 0.68\mathring{A}^{-1}$, and in a boundary
resistance $AR_{Co/Cr}^{\ast }(\mathrm{theor})\simeq 0.61\,f\Omega \,m^{2}$.
The Fermi momentum is satisfactory in the frame of the free electron model 
\cite{Ashk-Merm}. The boundary resistance is close enough to the
experimental value $AR_{Co/Cr}^{\ast }(\exp )\simeq 0.48\,f\Omega \,m^{2}$ 
\cite{Vouille1}. We expect that better matching of the band structures of 
cobalt and chromium, both belonging to the iron-group metals, would results 
in a weaker influence of the real band structure on the boundary resistance.

A discrepancy with experiment in an absolute value of boundary 
resistance of about 20-45\% seems not catastrophic because of 
the following reasons. First, the trial choice of 
$p_{\uparrow }^{F}=1.0\mathring{A}^{-1}$ was not optimized. 
Second, we used the free \textit{s}-electron model as a 
background for the theory. One may expect that for an interface 
between a metal with predominantly \textit{d}-electron conduction 
band (iron group) and an \textit{s}-electron metal (Cu, Ag), 
reduced overlapping and symmetry mismatch may noticeably increase 
the boundary resistance. Third, Garc\'{\i}a and Stoll have shown 
(Fig. 2 of Ref. \cite{G-S}) that the interface roughness 
also increases boundary resistance. This increase is estimated 
below 20-60\% for different models and magnitudes of the interface 
roughness, and reasonable differences in the Fermi momenta of contacting 
metals. The interface roughness is much less important for the case of 
CPP transport, considered in this paper, than for current-in-plane 
(CIP) transport \cite{Garcia-pc}. Finally, spin-reversal at the 
interface opposes the both previously considered processes, 
decreasing the boundary resistance. The reversal of the electron 
spin by the spin-orbit interaction is always expected upon refraction 
of the electron wave or scattering on roughness at the interface 
between two metals. Quantitative analysis of the competition between 
the above minor mechanisms of boundary resistance is beyond the scope 
of the paper. However, it is our expectation that the key quantity, 
spin asymmetry of the boundary resistance $\gamma_{F/N}$, is only weakly 
dependent on band structures matching, interface roughness, 
spin reversal \textit{etc}. because of considerable cancellations in 
the dimensionless ratio, Eq. (\ref{eq21}).

Our trial evaluations show that the experimental data for the spin asymmetry
of boundary resistance and the absolute value of boundary resistance in 
$Co/Cu$, $Co/Ag$ and $Co/Cr$ multilayers can be consistently described with
the use of the spin polarization parameter for the conduction band of cobalt 
$\delta \simeq 0.6$. At the level of the experimental accuracy and 
completeness of the theory the estimated value of $\delta $ is identical to 
$\delta\simeq 0.57$, which we have estimated \cite{Tag} from experiments of 
Garc\'{\i}a {\it et al.} \cite{Gar-Co} on magnetoresistance of cobalt 
nanocontacts. Experiments on Andreev spectroscopy cite similar values of 
$\delta $ for cobalt \cite{Soulen99}. Thus, the spin asymmetry of the 
boundary resistance in combination with the absolute value of the 
boundary resistance can be used for estimations of the 
spin-polarization parameter $\delta$ of the conduction 
band of ferromagnetic metals.

\ack

The work was supported by EC via the Contract No NMP4-CT-2003-505282, and 
by RFBR via projects No 03-02-17432 and No 03-02-17656. L R T acknowledges
numerous illuminating discussions with Professor Nicolas Garc\'{\i}a.


\begin{center}
\textbf{Figure captions}
\end{center}

Fig. 1. Dependence of the spin-asymmetry of boundary resistance $\gamma
_{F/N}$ (top field) and the renormalized resistance of the interface 
$AR_{F/N}^{\ast }$ (bottom field) on the Fermi momentum of the 
non-magnetic metal for the case 
($p_{\uparrow }^{F}>p_{\downarrow }^{F}>p^{N}$).

Fig. 2. The same as in Figure 1, but for the case ($p_{\uparrow
}^{F}>p^{N}>p_{\downarrow }^{F}$).

Fig. 3. The same as in Figure 1, but for the case ($p^{N}>p_{\uparrow
}^{F}>p_{\downarrow }^{F}$).

\end{document}